\documentclass[11pt]{cernrep}
\usepackage{graphicx}
\usepackage{cite,./mcite}
\begin{document}
\title{The Les Houches Accord PDFs (LHAPDF) and LHAGLUE}

\author{M R Whalley$^\dagger$, D Bourilkov$^\ddagger$ and R C Group$^\ddagger$}
\institute{
University of Durham$^\dagger$\\
Durham, DH1 3LE, UK\\
and\\
University of Florida$^\ddagger$\\
Gainesville, FL 32611,USA
}

\maketitle
\begin{abstract}
We describe the development of the LHAPDF library from its initial implementation 
following the Les Houches meeting in 2001 to its present state as a functional
replacement for PDFLIB. Brief details are given of how to install and use the 
library together with the PDF sets available. We also describe LHAGLUE,
an add-on PDFLIB look-a-like interface to LHAPDF, which  facilitates using
LHAPDF with existing Monte Carlo generators such as PYTHIA and HERWIG.    
\end{abstract}

\section{Working group summary}
\label{sec:lhapdf}
\subsection{LHAPDF - Introduction}

Parton Density Functions (PDFs), which describe the partonic content of hadrons,
need to be well understood and of sufficiently 
high precision if theoretical predictions are to match the experimental 
accuracies expected from future LHC data.  
These PDFs, which are produced by several different groups
(e.g. MRST, CTEQ, Alekhin and more recently ZEUS and H1), are 
derived from fitting deep inelastic and related hard scattering data
using parameterisations at 
low $Q_0^2$ ($\approx$ 1--7 (GeV/c)$^2$) and evolving these to higher 
$Q^2$. 
These PDFs are typically presented as grids in 
$x$-$Q^2$ with suitable interpolation codes provided by the PDF authors. 
The CERN PDFLIB library~\cite{pdflib} has to date provided a widely used 
standard 
FORTRAN interface to these PDFs with the interpolation grids
built into the PDFLIB code itself. However, it is realised
that PDFLIB would be increasingly unable to meet the needs of the new 
generation of PDFs which often involve large numbers of sets ($\approx$20--40)
des
cribing the uncertainties on the individual partons from  variations in the fitted parameters.  
As a consequence of this, at the Les Houches meeting in 2001~\cite{giele:2002hx},
the beginnings of a new interface were conceived --- the so-call \lq\lq Les Houches Accord PDF\rq\rq
 --- LHAPDF. This has further been developed over the course of the HERA-LHC workshop
incorporating many new features to enable it to replace PDFLIB as the standard tool to use.
The development is briefly described in this writeup together with LHAGLUE, an interface
to LHAPDF, 
which provides PDF access using almost identical calling routines as PDFLIB. 

\subsection{LHAPDF - Development during the Workshop} 

In its initial incarnation (Version 1), LHAPDF had two important features which distinguished it from the 
methods used by PDFLIB in handling PDFs.

Firstly the PDFs are defined by the analytical formulae used in the original fitting procedures, with external files of parameters,
which describe
the momentum $x$ distributions of the partons at the relevant $Q_0^2$. Evolution codes within
LHAPDF then produce the PDF at any desired $Q^2$ at the users request. 
At present LHAPDF provides access to two evolution codes, EVLCTEQ for the CTEQ distributions
and QCDNUM 16.12 \cite{qcdnum} for the other PDF sets.
This represents a radical difference from the existing methods used by the PDF authors to present their 
distributions where large grid files and interpolation routines are the norm. In PDFLIB these
interpolation codes and grids are essentially compiled into a single FORTRAN library. 
The advantage of the LHAPDF method is that the compiled code is separate
from  the parameter files, which are typically small. Thus to add new PDF sets does not necessarily need the code to be 
recompiled and the library rebuilt.  

Secondly, the concept is introduced of a \lq\lq set" being a related collection of PDFs (e.g. an error set)
all of which are accessible to the program after initialisation of that set.
This allows LHAPDF to handle the multi-set \lq\lq error" PDFs produced in recent years 
which give predicted uncertainties to the PDF values. All the PDFs in a set are initialised together 
and are therefore available to the user.   

V1 was written by Walter Giele of Fermilab who in 2002 released a working 
version which could be downloaded from a web-site together with the parameter files for a 
limited number of PDF sets.  There was also a manual and example files.  One of the present authors
MRW became involved and took over maintenance and development of LHAPDF in March 2003.
The limitations of the idealised situation in V1 with respect to making LHAPDF a replacement
tool for PDFLIB soon became apparent.  

The primary problem was that V1 contained only a limited number of PDF sets and, since the method was
reliant on the $x$ parameterisations at  $Q_0^2$ being available, it would be virtually
impossible to include many of the older sets which are still needed for comparisons.  
A second and serious problem is the compute time taken in the initialisation phase of the 
individual members of a PDF set (i.e. calling the routine InitPDF described later). 
This can take in the region of 2 seconds per call on a 1GHz machine and is therefore unacceptable in
the situation of a program which makes repeated use of the different members.~\footnote{A third problem reported
at the workshop concerning small differences (up to $\approx0.5\%$) between the PDFs produced by LHAPDF 
for MRST and the authors' code directly is now believed to be due to slight mismatches of grid
boundaries at the heavy quark thresholds and will be corrected in future MRST grids.}

A solution introduced in LHAPDF Version 2, which helps to solve the above problems, was to include
the option to make the original grid files and interpolation codes available in LHAPDF {\bf in addition}
to the V1 method of parameter files and \lq\lq on-the-fly" evolution. For some PDF sets both methods would
be available and for others only the latter.  The operation of the program was made  
identical for both methods with the content of the input file (with extension \lq\lq .LHpdf" for the former and 
\lq\lq .LHgrid" for the latter) dictating which is used.  Not only does this allow all the older PDF sets to 
be included but also there is no time penalty in changing
between members of the same set since all are loaded in the initialisation phase.
LHAPDF V2 was released in March 2004 including many of the older PDF sets as well as some new ones.       

LHAPDF Version 3 was released in September 2004 and, as well incorporating more older and some new PDF sets 
(e.g. ZEUS and H1), it also included the code for LHAGLUE, a newly developing add-on interface to LHAPDF
which provides PDFLIB look-alike access.  In addition to having subroutine calls identical to those in
PDFLIB it also incorporates a PDF numbering scheme to simplify usage. It should be noted however that, because of 
the greatly increased number of new PDF sets, it was not possible to follow the original numbering scheme of 
PDFLIB and a new one was devised.  
This is described in more detail in Section \ref{lhaglue_section}.

The major feature of Version 4, which was released in March 2005, was the incorporation of the photon and pion PDFs.
All the photon and pion PDFs that were implemented in PDFLIB were put into LHAPDF using identical code and
using  the \lq\lq .LHgrid" method. The LHAGLUE numbering scheme in these cases more closely resembles that of
PDFLIB than it does for the protons.

In addition in V4 there were new proton PDFs (MRST2004 and an updated Alekhin's a02m), a new simpler file structure
with all the source files being in a single \lq\lq src" directory, some code changes to 
incorporate access to $\Lambda_{4/5}^{QCD}$ and a  more rigorous implementation of the $\alpha_s$ evolution as being exactly
that used by the PDF author. 

All the LHAPDF and LHAGLUE data and code, in addition to being made available on the new web site 
(http://hepforge.cedar.ac.uk/lhapdf/), 
is also included in the GENSER subproject of the LHC Computing Grid.

\subsection{LHAPDF - Development after the Workshop} 

Since the last HERA-LHC meeting there has been one minor release of LHAPDF (Version 4.1 in August 2005).
In this version the installation method has changed to be more  standard with the 
\lq\lq configure; make; make install" sequence familiar to many and also a small amount
of code has been altered to be more compliant with proprietary FORTRAN 95 compilers. As mentioned in the previous section
the web site for public access to  
LHAPDF from which the source code can be obtained has changed.
Since this is the current and most recent version we assume V4.1 in the following referring to earlier versions where
necessary.

\subsection{Using LHAPDF}

Once the code and PDF data sets have been downloaded from the relevant web site 
and installed following the instructions given therein, using LHAPDF is simply a 
matter of linking the compiled FORTRAN library {\bf libLHAPDF.a} to the users 
program. Table \ref{lhapdf-commands} lists the LHAPDF routines available to the user,
which are of three types:
\begin{itemize}
\item Initialisation (selecting the required PDF set and its member)
\item Evolution (producing the momentum density functions ({\sl f}) for the partons at selected {\sl x} and {\sl Q})
\item Information (displaying for example $\alpha_s$, descriptions, etc.)
\end{itemize}

\begin{table}[h]
\begin{center}
\begin{tabular}{|l|l|}
\hline
Command & Description  \\
\hline

\bf call InitPDFset(\sl name\ ) &  Initialises the PDF set to use.  \\ 
\bf call InitPDF(\sl member\ ) &  Selects the member from the above PDF set.  \\ 
\hline
\bf call evolvePDF(\sl x,Q,f\ ) & Returns the momentum density function, {\sl f}$(x,Q$), for protons or pions.\\ 
\bf call evolvePDFp(\sl x,Q,P2,ip2,f\ ) & Returns the momentum density function  for photons.~\footnotemark.\\ 
\hline
\bf call numberPDF(\sl num\ ) & Returns the number ({\sl num}) of PDF members in the set.\\
\bf call GetDesc(\ ) & Prints a description of the PDF set.\\
\bf  alphasPDF({\sl Q\ }) & Function giving the value of $\alpha_s$ at {\sl Q} GeV.\\
\bf call GetLam4({\sl mem,qcdl4\ }) & Returns the value of $\Lambda_4^{QCD}$ for the specific member. \\
\bf call GetLam5({\sl mem,qcdl5\ }) & Returns the value of $\Lambda_5^{QCD}$ for the specific member.\\
\bf call GetOrderPDF({\sl order\ }) & Returns the order of the PDF evolution.\\
\bf call GetOrderAs({\sl order\ }) & Returns the order of the evolution of $\alpha_s$ . \\
\bf call GetRenFac({\sl muf\ }) & Returns the renormalisation factor.\\
\bf call GetQmass({\sl nf,mass\ }) & Returns the mass of the parton of flavour {\sl nf}. \\
\bf call GetThreshold({\sl nf,Q\ }) & Returns the threshold value for parton of flavour {\sl nf}.\\
\bf call GetNf({\sl nfmax\ }) & Returns the number of flavours.\\
\hline
\end{tabular}
\caption{LHAPDF commands}\label{lhapdf-commands}

\end{center}
\end{table}

\footnotetext{In {\bf evolvePDFp} $P2$ is the vitruality of the photon in GeV\ $^2$, which should by 0 for an on-shell photon, and {\sl ip2} 
is the parameter to evaluate the off-shell anomalous component. See the PDFLIB manual~\cite{pdflib} for details.}

The evolution commands utilise a double precision array {\sl f(-6:6)} 
where the arguments range from -6 to +6 for the different (anti)partons as shown in 
Table~\ref{flavour-scheme} below. 
 
\begin{table}[h]
\begin{center}
\begin{tabular}{
|p{1.0cm}
*{13}{|p{0.4cm}}
|}
\hline
parton &
$\bar{t}$ & 
$\bar{b}$ & 
$\bar{c}$ & 
$\bar{d}$ & 
$\bar{u}$ & 
$\bar{d}$ & 

g &
d &
u &
s &
c &
b &
t \\
\hline
{\sl n} & -6 & -5 & -4 & -3 & -2 & -1 & 0 & 1 & 2 & 3 & 4 & 5 & 6 \\
\hline
\end{tabular}
\caption{The flavour enumeration scheme used for {\sl f(n)} in LHAPDF}
\label{flavour-scheme}
\end{center}
\end{table}

Specifying the location of the PDF sets in the code should be especially mentioned at 
this point.  The argument ({\sl name}) in {\bf InitPDFset} should specify
the complete path (or at least to a symbolic link to this path). From version 4.1 
onwards, however, a new routine {\bf InitPDFsetByName} can be used in which only the name of the PDF 
set need by specified.  This works in conjunction with the script "lhapdf-config" 
which is generated at the configure stage of the installation which  provides
the correct path to the PDF sets. The location of this script must therefore be in 
the users execution path. Tables \ref{lhapdf-sets1} and \ref{lhapdf-sets2} list the complete range of PDF set
available. The equivalent numbers to use in LHAGLUE, as described in the next section, 
are also listed in these tables.

\begin{table}[h]
\begin{center}
\begin{tabular}{|l|l|l|c|l|}
\hline
Ref & Prefix & Suffix (\# of sets) & type & LHAGLUE numbers  \\
\hline
\hline
\multicolumn{5}{|c|}{\bf Proton PDFs}     \\
\cite{Alekhin:2000ch} & alekhin\_  & 100 (100), 1000 (1000)           & p   & {\bf 40100-200}, {\bf 41000-1999}  \\
\cite{Alekhin:2002fv} & a02m\_     & lo (17), nlo (17), nnlo (17) ) ) & g   & 40350-67, 40450-67, 40550-67\\
\cite{Botje:1999dj}   & botje\_    & 100 (100),1000 (1000)            & p   & {\bf 50100-200}, {\bf 51000-1999}   \\
\cite{Stump:2003yu}   & cteq       & 61 (41)                          & p,g & {\bf 10100-40}, 10150-90 \\
\cite{Pumplin:2002vw} & cteq       &6 (41)                            & p,g & {\bf 10000-40}, 10050-90 \\
                      & cteq       & 6m, 6l, 6ll                      & p   & {\bf 10040, 10041, 10042} \\
\cite{Lai:1999wy}     & cteq       & 5m, 5m1, 5d, 5l                  & g   & 19050, 19051, 19060, 19070 \\
\cite{Lai:1996mg}     & cteq       & 4m, 4d, 4l                       & g   & 19150,  19160, 19170  \\
\cite{Giele:2001mr}   & fermi2002\_& 100 (100), 1000 (1000)           & p   & {\bf 30100-200}, {\bf 31000-2000}  \\
\cite{Gluck:1998xa}   & GRV98      &lo, nlo(2)                        & g   & 80060, 80050-1  \\ 
\cite{Adloff:2003uh}  & H12000     & msE (21), disE (21), loE (21) & g   & 70050-70, 70150-70, 70250-70 \\  
\cite{Martin:2004dh}  & MRST2004   & nlo                              & p,g & {\bf 20400}, 20450  \\
                      & MRST2004   &nnlo                              & g   & 20470 \\
\cite{Martin:2003tt}  & MRST2003   &cnlo                              & p,g & {\bf 20300}, 20350  \\
                      & MRST2003   &cnnlo                             & g   & 20370 \\
\cite{Martin:2002aw}  & MRST2002   & nlo (2)                          & p,g & {\bf 20200}, 20250 \\
                      & MRST2002   &nnlo                              & g   & 20270  \\
                      & MRST2001   & E (31)                           & p,g & {\bf 20100-130}, 20150-180  \\
\cite{Martin:2001es}  & MRST2001   & nlo(4)                           & p,g & {\bf 20000-4}, 20500-4  \\
                      & MRST2001   &lo, nnlo                          & g   & 20060, 20070  \\
\cite{Martin:1998sq}  & MRST98     & (3)                              & p   & {\bf 29000-3}  \\
                      & MRST98     & lo (5), nlo (5) dis (5), ht      & g   & 29040-5, 29050-5,29060-5,29070-5\\
\cite{Chekanov:2002pv}& ZEUS2002\_ & TR (23), FF (23), ZM (23)        & p   & {\bf 60000-22}, {\bf 60100-22}, {\bf 60200-22} \\
\cite{Chekanov:2005nn}& ZEUS2005\_ & ZJ (23)                          & p   & {\bf 60300-22} \\
\hline
\multicolumn{5}{|l|}{\bf Notes:} \\
\multicolumn{1}{|c}{} & \multicolumn{1}{r}{LHAPDF $\rightarrow$} 
                      & \multicolumn{1}{l}{PrefixSuffix.LHpdf  (type p).}
		      & \multicolumn{2}{l|}{Where both p and g are present (p,g) then} \\
\multicolumn{1}{|c}{} & \multicolumn{1}{r}{filename $\rightarrow$} 
                      & \multicolumn{1}{l}{PrefixSuffix.LHgrid  (type g).} 
		      & \multicolumn{2}{l|}{the user has the choice of either.} \\[0.2cm]
\multicolumn{1}{|c}{} & \multicolumn{4}{l|}{LHAGLUE numbers in {\bf bold} are the type p (.LHpdf) sets.} \\
\hline
\end{tabular}
\caption{The Proton PDF sets available in LHAPDF.}\label{lhapdf-sets1}
\end{center}
\end{table}

\begin{table}[h]
\begin{center}
\begin{tabular}{|l|l|l|l|l|l|}
\hline
Prefix & Suffix  & LHAGLUE numbers  & Prefix & Suffix  & LHAGLUE numbers \\
\hline
\hline
\multicolumn{3}{|c|}{\bf Pion PDFs} &  \multicolumn{3}{|c|}{\bf Photon PDFs}      \\
 OWPI       & (2)             & 211-12    & DOG        & 0, 1		 & 311, 312    \\
 SMRSPI     & (3)	      & 231-3     & DGG        & (4)		 & 321-4       \\
 GRVPI      & 0, 1	      & 251, 252  & LACG       & (4)		 & 331-4       \\
 ABFKWPI    & (3)             & 261-3     & GSG        & 0 (2), 1	 & 341-2, 343  \\
\cline{1-3}
\multicolumn{3}{|l|}{All filenames are PrefixSuffix.LHgrid} 	      & GSG96      & 0, 1		 & 344, 345    \\ 
\cline{1-3}
\multicolumn{3}{|l|}{The nomenclature used here is}	      & GRVG	 & 0 (2), 1 (2)    & 351-2, 353-4 \\
\multicolumn{3}{|l|}{essentially the same as in PDFLIB and}	      & ACFGPG	 & (3)  	   & 361-3	 \\
\multicolumn{3}{|l|}{the relevant publication references}	      & WHITG	 & (6)  	   & 381-6	 \\
\multicolumn{3}{|l|}{can be found in the PDFLIB manual \cite{pdflib}.}	      & SASG	 & (8)  	   & 391-8	 \\
\hline
\end{tabular}
\caption{The Pion and Photon PDF sets available in LHAPDF.}\label{lhapdf-sets2}
\end{center}
\end{table}

\subsection{LHAGLUE}
\label{lhaglue_section}
      The LHAGLUE interface~\cite{Bourilkov:2003kk} to LHAPDF is designed along the lines of the existing interface
from PYTHIA to PDFLIB.~\footnote{DB
would like to thank T.~Sj\"ostrand and S.~Mrenna for discussions on this
topic.}.
For both HERWIG and PYTHIA the existing 'hooks' for
PDFLIB have been utilised for the LHAGLUE interface.
This makes it possible to link it exactly like PDFLIB with no further changes
to PYTHIA's or HERWIG's source code needing to be implemented.
 
The interface contains three subroutines (similar to PDFLIB) and can be used
seamlessly by Monte Carlo generators interfaced to PDFLIB or in
standalone mode. These are described in Table~\ref{LHAGLUE-commands}.
In addition any of the LHAPDF routines, {\bf except} the initialisation routines
{\bf InitPDFset} and {\bf InitPDF}, described in Table~\ref{lhapdf-commands}, can also be used, for
example to return the value of the strong coupling constant $\alpha_s$
({\bf alphasPDF($Q$)}), or to print the file description ({\bf call GetDesc()}).

There are also several CONTROL switches specified through
the 20 element character array LHAPARM and COMMON blocks  
which determine how the interface operates.

\begin{itemize}
\item Location of the LHAPDF library of PDFs (pathname):\hfill\break
      From version LHAPDF v4.1 onwards, and the LHAGLUE routines distributed 
      with it, the location of the PDFsets data files is set automatically 
      using the "lhapdf-config" script as described in the previous section,
      provided that the prescribed installation 
      instructions have been used.\hfill\break
      For previous versions (4.0 and earlier)  the common block 
      {\bf COMMON/LHAPDFC/LHAPATH} is used where {\bf LHAPATH} is a 
      {\bf character*132} variable containing the full path to the PDF sets.
      The default path is subdir 'PDFsets' of the current directory. 
\item Statistics on under/over-flow requests for PDFs outside their 
      validity ranges in {\sl x} and $Q^2$.\hfill\break
      a) {\bf LHAPARM(16) .EQ. \lq NOSTAT'} $\rightarrow$ No statistics (faster)\hfill\break
      b) {\bf LHAPARM(16) .NE. \lq NOSTAT'} $\rightarrow$ Default: collect statistics\hfill\break
      c) {\bf call PDFSTA} at the end to print out statistics.\hfill\break
\item Option to use the values for $\alpha_s$ as computed by LHAPDF
      in the Monte Carlo generator as well in order to ensure uniform
      $\alpha_s$ values throughout a run \hfill\break 
      a) {\bf LHAPARM(17) .EQ. \lq LHAPDF'} $\rightarrow$ Use $\alpha_s$ from LHAPDF\hfill\break
      b) {\bf LHAPARM(17) .NE. \lq LHAPDF'}$\rightarrow$  Default (same as LHAPDF V1/V3)\hfill\break
\item Extrapolation of PDFs outside the LHAPDF validity range given by {\sl x}$_{min/max}$ and {\sl Q}$^2_{min/max}$.\hfill\break
      a) Default $\rightarrow$  PDFs \lq\lq frozen" at the boundaries.\hfill\break
      b) {\bf LHAPARM(18) .EQ. \lq EXTRAPOLATE'} $\rightarrow$ Extrapolate PDFs at own risk\hfill\break
      \item Printout of initialisation information in PDFSET (by default)\hfill\break
      a) {\bf LHAPARM(19) .EQ. \lq SILENT'}  $\rightarrow$ No printout (silent mode).\hfill\break
      b) {\bf LHAPARM(19) .EQ. \lq LOWKEY'} $\rightarrow$ Print 5 times (almost silent).\hfill\break
\item Double Precision values of $\Lambda_{4/5}^{QCD}$  applicable to the selected PDF are available (as read-only) 
in the COMMON block:\ {\bf COMMON/W50512/QCDL4,QCDL5} $\rightarrow$ as in PDFLIB.\hfill\break 
\end{itemize}

\begin{table}[ht]
\begin{center}
\begin{tabular}{|p{2in}|l|}
\hline

Command & Description \\[0.1cm]
\hline
\hline 
{\bf call PDFSET(\sl parm,value)}
 & For initialisation (called once) where PARM and VALUE \\
 & are LOCAL arrays in the calling program specified as  \\
 &     {\bf  \ \ \ CHARACTER*20 PARM(20)} \\
 &     {\bf  \ \ \ DOUBLE PRECISION VALUE(20)} \\[0.1cm]
\hline
\multicolumn{2}{|l|}{\bf call STRUCTM(\sl X,Q,UPV,DNV,USEA,DSEA,STR,CHM,BOT,TOP,GLU)}  \\
 & For the proton (and pion) PDFs: where X and Q are the \\
 & input kinematic variables and the rest are the output\\
 & PDF of the valence and sea quarks and the gluon.  \\[0.1cm] 
 \hline
\multicolumn{2}{|l|}{\bf call STRUCTP(\sl 
X,Q2,P2,IP2,UPV,DNV,USEA,DSEA,STR,CHM,BOT,TOP,GLU)} \\
 & For the photon PDFs: as above with the additional input\\ 
 & variables P2 and IP2~\footnotemark[2]. \\[0.1cm]
\hline
\end{tabular}
\caption{LHAGLUE commands}\label{LHAGLUE-commands}
\end{center}
\end{table}

The LHAGLUE interface can be invoked in one of 3 ways, Standalone, PYTHIA or 
 HERWIG, depending on the value of {\sl parm(1)} when calling {\bf PDFSET({\sl parm,value)}}.

\begin{itemize}
\item{\bf Standalone} mode \hfill\break
      PARM(1)= \lq DEFAULT' \hfill\break
      VALUE(1) = \lq\lq{\sl PDF number}" 

%
%
%
\item {\bf PYTHIA mode}\hfill\break
    PARM(1) = 'NPTYPE' $\leftarrow$ set automatically in PYTHIA \hfill\break
    In this case the user must supply MSTP(51) and MSTP(52) in the PYTHIA common block\hfill\break
    {\bf COMMON/PYPARS/MSTP(200),PARP(200),MSTI(200).PARI(200)}\hfill\break
    MSTP(52) = 2 $\leftarrow$ to use an external PDF library\hfill\break
    MSTP(51)= \lq\lq{\sl PDF number}"
\item {\bf HERWIG mode}\hfill\break
    PARM(1) = 'HWLHAPDF'$\leftarrow$ set by the user.\hfill\break
    In this case one sets for the beam and target particles separately\hfill\break
      AUTPDF(1) = 'HWLHAPDF'\hfill\break
      AUTPDF(2) = 'HWLHAPDF'\hfill\break
      MODPDF(1) = \lq\lq{\sl PDF number}"\hfill\break
      MODPDF(2) = \lq\lq{\sl PDF number}"\hfill\break
Note that HERWIG specifies the\lq\lq{\sl PDF number}" for each of the colliding particles
separately and care should be taken that the same PDF members are used when appropriate.
\end{itemize}

\noindent The user then simply links their own standalone code, or the HERWIG/PYTHIA main
program and the HERWIG/PYTHIA code~\footnote{It is important when starting with a 
fresh PYTHIA or HERWIG download the user
must first rename the 'dummy' subroutines {\bf STRUCTM}, {\bf STRUCTP} and {\bf PDFSET} in the
PYTHIA/ HERWIG source codes exactly as if one were to link to PDFLIB.},
with  the LHAPDF
library {\bf libLHAPDF.a} making sure the \lq PDFsets' directory is
specified as described above. 

The LHAGLUE interface has been tested extensively at TEVATRON and LHC energies
for the proton PDFs and with HERA examples for the photon PDFs. Results with new 
and legacy PDF sets, using LHAPDF, PDFLIB or internal implementations in the
Monte Carlo generators, and comparing cross sections produced with PYTHIA and
HERWIG, give us confidence in the consistency of the LHAGLUE interface and the
underlying LHAPDF library~\cite{noteinprep}.

\subsection{Summary and Future Development}

Both LHAPDF and the interface LHAGLUE have been developed over the period of 
the Workshop to a point where they can now be used as a serious 
replacement for PDFLIB.  Indeed, except for the PDF authors' own code, they are
the only place to obtain the latest PDFs.  There is however still considerable
development in progress and the latest PDF sets will be incorporated as and
when they become available. One major development area is to include the possibility
of having more than one PDF set initialised concurrently.  This may be necessary in 
 interactions between different beam and target
particles types and also including  photon and pion PDFs.
This will be the aim of the next LHAPDF release.

\section*{Acknowledgements}
MRW wishes to thank the UK PPARC for support from grant PP/B500590/1.
DB wishes to thank the USA National Science Foundation
for support under grants NSF ITR-0086044 and NSF PHY-0122557. \\

\bibliographystyle{heralhc} 
{\raggedright
\bibliography{lhapdf}
}
\end{document}